\def\be{\begin{equation}}
\def\ee{\end{equation}}
\def\ba{\begin{array}}
\def\ea{\end{array}}

\documentclass[prl,showpacs,twocolumn,amsmath]{revtex4}
\usepackage{amsfonts}
\usepackage{mathrsfs}
\usepackage{amssymb}
\usepackage{pifont}
\usepackage{epsfig,subfigure,dsfont,amsthm,amsbsy,mathrsfs,amscd}
\def\qed{\leavevmode\unskip\penalty9999 \hbox{}\nobreak\hfill
     \quad\hbox{\leavevmode  \hbox to.77778em{%
               \hfil\vrule   \vbox to.675em%
               {\hrule width.6em\vfil\hrule}\vrule\hfil}}
     \par\vskip3pt}
\usepackage{leftidx}

\input amssym.def

\begin{document}
\title{\large\bf Local projective measurement-enhanced quantum battery capacity}
\author{ Tinggui Zhang$^{1, \dag}$, Hong Yang$^{2}$ and Shao-Ming Fei$^{3}$}
\affiliation{ ${1}$ School of Mathematics and Statistics, Hainan Normal University, Haikou, 571158, China \\
$2$ School of Physics and Electronic Engineering, Hainan Normal University, Haikou, 571158, China \\
$3$ School of Mathematical Sciences, Capital Normal University, Beijing 100048, China \\
$^{\dag}$ Correspondence to tinggui333@163.com}

\bigskip
\bigskip

\begin{abstract}
Quantum battery is of significant potential applications in future
industry and daily life. The battery capacity is an important
indicator of a battery. How to improve the capacity of quantum
batteries is of importance. We consider quantum batteries given by
bipartite quantum systems, and study the enhancement of the battery
capacity under local projective measurements on a subsystem of the
quantum state. By using the two-qubit Bell diagonal states and the
X-type states as examples, we show that the quantum battery capacity
with respect to the whole system or the subsystem could be improved
by local projective measurements. Our theoretical analysis will
provide new ideas for the experimental development of quantum
battery.

\end{abstract}

 \maketitle

\section{Introduction}
As an energy storage system, the battery
plays significant roles in both industry and daily life.
With the development of quantum technology and information science,
the quantum battery has emerged and been expected to balance the advantages
of small size, large capacity, portability, rapid charging etc. R. Alicki
and M. Fannes first formally introduced the idea of quantum
batteries in an informational theoretic context by characterizing
the maximum amount of energy that can be extracted from a quantum
system under unitary operations \cite{ramf}. Since then there have been extensive theoretical \cite{mkmh,mkmp,egnm,cdjo,fffl,gjfm,gdav,gfrz,dmgv,tjkm,agnj,dgmp,gmav,rali,ammd,stas,sfff,
dgdm,mtjj,stsa,ffqz,ksus,srvg,yyxs,sfss,khgw,sgas,fmaj,jcrj,srsv,fhja,rbgp,khzh,tlsa,
hsqx,fyyj,fyja,rgdv,khhg,jsjj,xymr,srsv2,fdfy,ksus2,tscg,bham,asov,paku,fsjm,dmam,hhqk,pbfn} and experimental\cite{jkld,aprm,jjts,ndfa,jjis} studies on quantum batteries. For example, many theoretical protocols have been proposed, such as charging by entangling operations\cite{fffl,gjfm,gfrz,gmav,fmaj,hsqx}, charging with dissipative\cite{fmaj,jcrj}, charging collectively and in parallel\cite{dmgv,jcrj}. Many-body interaction and energy fluctuation were also explored\cite{tjkm,dgmp,yyxs,sfss,khgw,fyyj,fyja,asov}. The first model of a quantum battery that could be engineered in a solid-state architecture, was proposed by Ferraro et al.\cite{dmgv}. Many other concrete quantum battery models have been also proposed\cite{tjkm,dgdm,hhqk}. However, many aspects of the physics of quantum batteries remain unexplored, experimental work on quantum batteries is still in its infancy and a fully-operational proof of principle is yet to be demonstrated. For a review of quantum batteries, we refer to Ref.\cite{fsjm}.

An important quantitative indicator of the quality of quantum
batteries is the capacity (charging power, work storage, work
extraction) \cite{srsv,xymr}. The commonly
used quantity is the ergotropy functional \cite{srsv},
$$
E(\rho,H)=\max_{U\in\mathbb{U}(d)}\{Tr[\rho H]-Tr[U\rho U^{\dag}H]\},
$$
where $\rho$ is the quantum state of a $d$-dimensional quantum system $Q$ with Hamiltonian $H$, the maximum takes over the states under all unitary evolution
$\rho\rightarrow U\rho U^{\dag}$, with $\mathbb{U}(d)$ the set
of the unitary operators acting on $Q$.
More recently, in Ref. \cite{xymr} the authors provided a new definition of quantum
battery capacity,
\be\label{zyf1}
C(\rho,H):=\sum_{i=0}^{d-1}\epsilon_i(\lambda_i-\lambda_{d-1-i}),
\ee
where $\lambda_0\leq\lambda_1\leq\cdots \leq\lambda_{d-1}$ denote
the eigenvalues of the quantum state $\rho$ and
$\epsilon_0\leq\epsilon_1\leq\cdots \leq\epsilon_{d-1}$ denote the
eigenenergies of the Hamiltonian
$H=\sum_i\epsilon_i|\varepsilon_i\rangle\langle\varepsilon_i|$.
$C(\rho,H)$ is a Schur-convex functional of the quantum state
and does not change when the battery is unitarily charged or discharged.

Quantum measurements, particularly the projections onto a chosen
state, could change the transition rate of the measured system
\cite{dhmw}. Numerous measurement-based control schemes have been
applied to state purification \cite{hmjf,jhka}, information gain
\cite{jchm} and entropy production \cite{gtmp}. The measurement on
auxiliary systems may help to control the target systems, serving as
state-engineering scheme through a nonunitary procedure
\cite{cean,sran}. In deed, the idea of improving the desired effect
of the target system by measuring the auxiliary system has recently
been applied to quantum battery charging \cite{jsjj} and power
extraction \cite{paku}. In \cite{jsjj}, the authors established for
quantum battery a charging by-measurement framework based on rounds
of joint evolution and partial-projection. Starting from a thermal
state, the battery could also achieve a near-unit ratio of ergotropy
and energy through less than $N$ measurements, when a population
inversion is realized by measurements. In
\cite{paku} the authors focused on the energy
extraction instead of charging, that is, the protocol includes only
one measurement on the auxiliary qubit, but not a sequence of
measurements. Scenarios are investigated for both that
the initial state of the battery and the auxiliary is a product and an
entangled one. It is shown that the measurement-based protocol provides
significantly better energy extraction from a quantum battery than
that based on unitary operations.

Consider a bipartite system composed of the target quantum state $A$
and the auxiliary system $B$. Our question is how the quantum
measurements on the systems $B$ would affect the quantum battery
capacity of the system $A$ or the whole bipartite system. In this
paper, we formulate a unilateral measurement protocol and study such
local projective measurement-enhanced quantum battery capacity by
starting directly from general bipartite quantum states. In general, a complete quantum battery consists of a battery part and a charger part. Here, we only consider the battery part, that is, the whole system battery is composed by two-particle quantum state $\rho_{AB}$ and the system Hamiltonian $H_{AB}$. And the battery of the subsystem is composed by the reduced state $\rho_A$ and the Hamiltonian $H_A$ corresponding to the subsystem A. We use the definition of quantum battery capacity given by
Eq.(\ref{zyf1}). We find that for two-qubit Werner states, the local
measurements can enhance the overall or the sub-system battery
capacity.

\section{Measurement-based protocol}

Let $\rho_{AB}$ be an $m\otimes n$ bipartite state in Hilbert space
$\mathcal {H}_A\otimes\mathcal {H}_B$. Let
$H_A=\sum_{i=0}^{m-1}\epsilon_i^A|\varepsilon_i\rangle^A\langle\varepsilon_i|$
and
$H_B=\sum_{i=0}^{n-1}\epsilon_i^B|\varepsilon_i\rangle^B\langle\varepsilon_i|$
be the Hamiltonian associated with the systems A and B,
respectively. The Hamiltonian of the joint system is
$H_{AB}=H_A\otimes I_n + I_m \otimes H_B$, where $I_n$ ($I_m$)
denotes the $n\times n$ ($m\times m$) identity matrix. The battery
capacity $C(\rho_{AB},H_{AB})$ can be calculated by using
Eq.(\ref{zyf1}). Let
$\rho_A=\sum_{i=0}^{m-1}\lambda_i|\psi_i\rangle\langle\psi_i|$ be
the reduced state of the subsystem $A$. We have the battery capacity
with respect to the subsystem $A$,
$C(\rho_A,H_A)=\sum_{i=0}^{m-1}\epsilon_i^A(\lambda_i-\lambda_{d-1-i}).$

Let $\{B_k\}_{k=0}^{n-1}$ be a rank-1 local projective measurement on the subsystem $B$.
Conditioned on the measurement outcome $k$, the quantum state $\rho_{AB}$ changes to be
\be
\rho_{AB}^{k}=\frac{1}{P_k}(I_m\otimes B_k)\rho_{AB}(I_m\otimes
B_k)
\ee
with probability $P_k=Tr(I_m\otimes B_k)\rho_{AB}(I_m\otimes
B_k)$ for $k=0,1,2,\cdots, n-1$. The final state
$\rho_{AB}^{\prime}$ depends on the measurement basis
as well as the outcomes. Under average probability (the group of measurements selected and the corresponding outputs are equally distributed), the final state is given by
\be\label{zyf2}
\rho_{AB}^{\prime}=\frac{1}{n}\sum_{k=0}^{n-1}\rho_{AB}^{k}.
\ee
With arbitrary randomness probability, namely, the measurement that one chooses to use is according to the idea of freedom and the probability of different outputs of the same measurement is also different, one has
\be\label{zyf3}
\rho_{AB}^{\prime}=\sum_{k=0}^{n-1}\mu_k\rho_{AB}^{k},
\ee
where $\sum_{k=0}^{n-1}\mu_k=1$, $\mu_k \geq 0$.
From the spectral decomposition of the reduced density matrix
$\rho_A^{\prime}=Tr_B(\rho_{AB}^{\prime})
=\sum_{i=0}^{m-1}\lambda_i^{\prime}|\psi_i\rangle^{\prime}\langle\psi_i|^{\prime}$, we obtain the capacity associated with the subsystem $A$ after projective measurement, $C(\rho_A^{\prime},H_A)=\sum_{i=0}^{m-1}
\epsilon_i^A(\lambda_i^{\prime}-\lambda^{\prime}_{d-1-i})$, see Fig. 1. for the schematic diagram. In the following we focus on two particular classes of initial states.
\begin{figure}[ptb]
\includegraphics[width=0.45\textwidth]{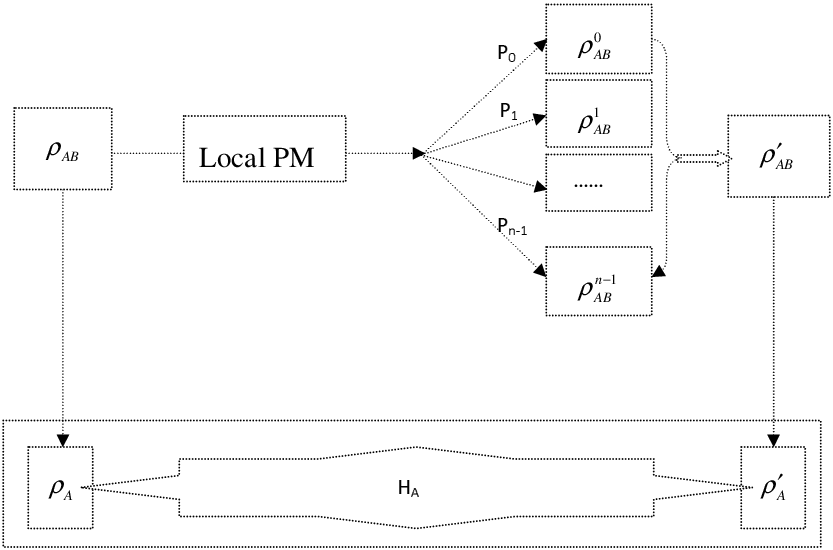}\caption{A initilaly shared bipartite quantum state $\rho_{AB}$ undergoes local projective measurements $\{B_k\}_{k=0}^{n-1}$ on the subsystem B. States $\rho_{AB}^k$ are obtained with probability $P_k$. From the final state $\rho_{AB}^{\prime}$ and its reduced density matrix $\rho_{A}^{\prime}$, we study the changes of the battery capacities before and after the measurement.}
\end{figure}

{\sf Two-qubit Bell-diagonal states}~~

We first consider the initial states to be the two-qubit Bell-diagonal ones:
$$
\rho_{AB}=\frac{1}{4}(I_2\otimes I_2+\sum_{i=1}^3c_i\sigma_i\otimes \sigma_i),
$$
where $\sigma_1,\sigma_2,\sigma_3$ are the standard
Pauli matrices, $c_i$ are real constants such that $\rho_{AB}$ is a well defined density
matrix. $\rho_{AB}$ have eigenvalues
$\lambda_0=(1-c_1-c_2-c_3)/4$, $\lambda_1=(1-c_1+c_2+c_3)/4$, $\lambda_2=(1+c_1-c_2+c_3)/4$ and $\lambda_3=(1+c_1+c_2-c_3)/4$. The coefficients $c_i$ are so chosen that
$\lambda_j\in[0,1]$ for $j=0,1,2,3$. For the convenience of future analysis,
we assume that $|c_1|\geq |c_2| \geq |c_3|$. We consider the following Hamiltonian for the two-qubit system,
\be\label{zyf4}
H_{AB}=\epsilon^A\sigma_3\otimes I_2+\epsilon^B I_2\otimes \sigma_3,
\ee
where $\epsilon^A \geq \epsilon^B \geq 0$. The eigenvalues of $H_{AB}$ are $-\epsilon^A-\epsilon^B$, $-\epsilon^A+\epsilon^B$,
$\epsilon^A-\epsilon^B$ and $\epsilon^A+\epsilon^B$ in ascending order.
Therefore, we have
$$
C(\rho_{AB},H_{AB})=(|c_1|+|c_2|)(\epsilon^A+\epsilon^B)+(|c_1|-|c_2|)(\epsilon^A-\epsilon^B).
$$
The reduced states of $\rho_{AB}$ are $\rho_A=I_2/2$ and $\rho_B=I_2/2$. Consequently, we have $C(\rho_A,H_A)=0$.

Consider local projective measurement
given by the computational base $\{|k\rangle\}$,
$\{\Pi_k=|k\rangle\langle k|,\ \ k=0,1\}$. The
measurement gives rise to the ensemble $\{\rho_{AB}^{k},P_k\}$. We have
\begin{eqnarray}\label{zyf5}
& & P_k\rho_{AB}^k=(I_2\otimes\Pi_k)\rho_{AB}(I_2\otimes \Pi_k)\nonumber\\
& =&\frac{1}{4}[I_2\otimes\Pi_k+\sum_{i=1}^3c_i\sigma_i\otimes
(\Pi_k\sigma_i\Pi_k)]\nonumber\\ & = &
\frac{1}{4}[I_2\otimes\Pi_k+(-1)^kc_3\sigma_3\otimes\Pi_k],
\end{eqnarray}
where we have used the relations $\Pi_k\sigma_j\Pi_k=0$ for $k=0,1,\
j=1,2$, $\Pi_0\sigma_3\Pi_0=\Pi_0$ and $\Pi_1\sigma_3\Pi_1=-\Pi_1$.
From Eq.(\ref{zyf5}) we get
$$\rho_{AB}^0=\frac{1}{2}\left(\begin{array}{cccc}
    1+c_3 & 0 & 0 & 0 \\
    0 & 0 & 0 & 0 \\
     0 & 0 & 1-c_3 & 0\\
      0 & 0 & 0 & 0
  \end{array}\right),
$$
$$\rho_{AB}^1=\frac{1}{2}\left(\begin{array}{cccc}
    0 & 0 & 0 & 0 \\
    0 &  1-c_3 & 0 & 0 \\
     0 & 0 & 0 & 0\\
      0 & 0 & 0 & 1+c_3
  \end{array}\right)
$$
as $P_0=P_1=\frac{1}{2}$.

By straightforward calculation we have the following observations.

(i) Taking the final state to be the one given by (\ref{zyf2}), we have
$\rho_{AB}^{\prime}=\frac{1}{2}\sum_{k=0}^{1}\rho_{AB}^{k}$. Then
$C(\rho_{AB}^{\prime},H_{AB})=2|c_3|\epsilon^A$ and
$C(\rho_A^{\prime},H_A)=0$. Therefore,
$F:=C(\rho_{AB}^{\prime},H_{AB})-C(\rho_{AB},H_{AB}) \leq 0$,
namely, the local measurement weakens the battery capacity of the
whole bipartite system, but does not change the battery capacity of
the subsystem $A$.

(ii) Taking the final state to be the one given by (\ref{zyf3}), we have
$\rho_{AB}^{\prime}=\mu_0\rho_{AB}^{0}+\mu_1\rho_{AB}^{1}$. Without
loss of generality, let $\mu_0 > \mu_1$. We obtain
$C(\rho_{AB}^{\prime},H_{AB})=(\mu_0-\mu_1+|c_3|)
(\epsilon^A+\epsilon^B)+(\mu_1-\mu_0+|c_3|)(\epsilon^A-\epsilon^B)$
and $C(\rho_A^{\prime},H_A)=2(\mu_0-\mu_1)\epsilon^A|c_3|$. In this
case, the local measurement could either reduce or increase the
battery capacity of the whole system, but always increase the
battery capacity of the subsystem $A$.

(iii) When $c_3=0$, $f:=C(\rho_A^{\prime},H_A)-C(\rho_A,H_A)$ is always zero.
However, for the $\rho_{AB}^{\prime}$ given in (ii), one has
$C(\rho_{AB}^{\prime},H_{AB})=2(\mu_0-\mu_1)\epsilon^B.$

The above observations show that the unilateral
measurement on subsystem $B$ may increase the battery capacity of
the subsystem $A$, but not always the entire system.

{\bf Example 1}: Let us consider a specific Bell-diagonal state with
$c_1=c_2=c_3=-a$, i.e., the two-qubit Werner state \cite{rfwe},
$$
\rho_{W}=a|\psi^-\rangle\langle\psi^-|+\frac{1-a}{4}I_4,
$$
where $|\psi^-\rangle=(|01\rangle-|10\rangle)/\sqrt{2}$ is the
maximally entangled state and $0 \leq a \leq 1$. $\rho_{W}$ is
separable when $a\leq \frac{1}{3}$ and entangled otherwise. From the
observation (ii), we see that after the projective measurement on
subsystem $B$, the battery capacity of the subsystem $A$ increases
as long as $a \neq 0$ and $\epsilon^A>0$, no matter if the original
state is separable or entangled. When $\mu_0-\mu_1 > a$, the
measurement on the subsystem $B$ can also increase the battery
capacity of the whole system. When $\mu_0-\mu_1 = a$, the battery
capacity of the whole system keeps unchanged. Therefore, for a
random probability combination Eq.(\ref{zyf3}), by choosing
appropriate parameters the local projective measurement may enhance
the battery capacity of the overall or the individual subsystem.

{\sf Two-qubit X-states}~~

We consider the initial states to be the two-qubit X-type ones,
\be\label{zyf6}
\rho_{X}=\left(\begin{array}{cccc}
    \rho_{11} & 0 & 0 & \rho_{14} \\
    0 & \rho_{22} & \rho_{23} & 0 \\
     0 & \rho_{32} & \rho_{33} & 0\\
      \rho_{41} & 0 & 0 & \rho_{44}
\end{array}\right),
\ee
where $\sum_{i=1}^4\rho_{ii}=1$, $\rho_{22}\rho_{33} \geq |\rho_{23}|^2$ and $\rho_{11}\rho_{44} \geq |\rho_{14}|^2$.
$\rho_{X}$ is entangled if and only if either $\rho_{11}\rho_{44} \leq
  |\rho_{23}|^2$ or $\rho_{22}\rho_{33} \leq
  |\rho_{14}|^2$ \cite{argv}.
The eigenvalues of the two-qubit X-states are given by
$$
\begin{array}{rcl}
\lambda_0&=&\frac{1}{2}[(\rho_{11}+\rho_{44})+\sqrt{(\rho_{11}-\rho_{44})^2+4|\rho_{14}|^2}],\\
\lambda_1&=&\frac{1}{2}[(\rho_{11}+\rho_{44})-\sqrt{(\rho_{11}-\rho_{44})^2+4|\rho_{14}|^2}],\\
\lambda_2&=&\frac{1}{2}[(\rho_{22}+\rho_{33})+\sqrt{(\rho_{22}-\rho_{33})^2+4|\rho_{23}|^2}],\\
\lambda_3&=&\frac{1}{2}[(\rho_{22}+\rho_{33})-\sqrt{(\rho_{22}-\rho_{33})^2+4|\rho_{23}|^2}].
\end{array}
$$
The reduced density matrix of the subsystem $A$,
\be\rho_{X}^A=\left(\begin{array}{cc}
    \rho_{11}+\rho_{22} & 0  \\
    0 & \rho_{33}+\rho_{44}
\end{array}\right),
\ee
has eigenvalues $\lambda_0^A=\rho_{11}+\rho_{22}$ and
$\lambda_1^A=\rho_{33}+\rho_{44}$. With respect to the Hamiltonian
(\ref{zyf4}), one has $C(\rho_X,H_{AB})$ and $C(\rho_X^A, H_{A})$.

$\rho_X$ can be rewritten as in Bloch representation,
$$
\rho_X=\frac{1}{4}(I_2\otimes I_2+a_3\sigma_3 \otimes I_2 +b_3
I_2\otimes \sigma_3+\sum_{i=1}^3c_i\sigma_i\otimes \sigma_i),
$$
where $a_3=\rho_{11}+\rho_{22}-(\rho_{33}+\rho_{44})$,
$b_3=\rho_{11}+\rho_{33}-(\rho_{22}+\rho_{44})$,
$c_1=2(\rho_{33}+\rho_{44})$, $c_2=2(\rho_{23}-\rho_{14})$ and
$c_3=\rho_{11}+\rho_{44}-(\rho_{22}+\rho_{33})$. After the
measurement $\Pi_k=|k\rangle\langle k|$ on subsystem $B$, we obtain
\begin{eqnarray}\label{zyf7} & &
P_k\rho_{X}^k=(I_2\otimes\Pi_k)\rho_{X}(I_2\otimes \Pi_k)\nonumber\\
&=&\frac{1}{4}[I_2\otimes\Pi_k+a_3\sigma_3\otimes\Pi_k
+b_3I_2\otimes(\Pi_k\sigma_3\Pi_k)\nonumber\\
&&+\sum_{i=1}^3c_i\sigma_i\otimes (\Pi_k\sigma_i\Pi_k)]\nonumber\\
&=& \frac{1}{4}[I_2\otimes\Pi_k+a_3(\sigma_3\otimes\Pi_k)\nonumber\\
&&+(-1)^kb_3(I_2\otimes\Pi_k)+ (-1)^kc_3(\sigma_3\otimes\Pi_k)].
\end{eqnarray}
Taking $k=0,1$ into $\Pi_k$, we obtain specific probabilities and
corresponding quantum states. Namely, $P_0=\frac{1+b_3}{2}$,
$P_1=\frac{1-b_3}{2}$ and
$$\rho_{X}^0=\frac{1}{2(1+b_3)}\left(\begin{array}{cccc}
    1+b_3+a_3+c_3 & 0 & 0 & 0 \\
    0 & 0 & 0 & 0 \\
     0 & 0 & 1+b_3-a_3-c_3 & 0\\
      0 & 0 & 0 & 0
  \end{array}\right),$$
$$\rho_{X}^1=\frac{1}{2(1-b_3)}\left(\begin{array}{cccc}
    0 & 0 & 0 & 0 \\
    0 & 1-b_3+a_3-c_3 & 0 & 0 \\
     0 & 0 & 0 & 0\\
      0 & 0 & 0 & 1-b_3-a_3+c_3
\end{array}\right).$$
From Eq.(\ref{zyf2}) or Eq.(\ref{zyf3}) we can investigate the
projective measurement impact on the battery capacity of subsystem
$A$ and the whole bipartite system.

{\bf Example 2}: Consider the following specific state,
$$
\rho_{AB}=\frac{1}{3}[(1-x)|00\rangle\langle00|
+2|\psi^+\rangle\langle\psi^+|+x|11\rangle\langle 11|],
$$
where $x\in[0,1/2]$ and
$|\psi^+\rangle=(|01\rangle+\langle10|)/\sqrt{2}$. $\rho_{AB}$ has
eigenvalues $\lambda_0=0$, $\lambda_1=x/3$, $\lambda_2=(1-x)/3$ and
$\lambda_3=2/3$. The reduced density matrix $\rho_A$ has eigenvalues
$\delta_0=(1+x)/3$ and $\delta_1=(2-x)/3$. Therefore,
$C(\rho_{AB},H_{AB})=\frac{2}{3}(\epsilon^A+\epsilon^B)
+\frac{1-2x}{3}(\epsilon^A-\epsilon^B)$ and
$C(\rho_{A},H_{A})=2\frac{1-2x}{3}\epsilon^A$. Adopting
Eq.(\ref{zyf2}) for $\rho_{AB}^{\prime}$, we obtain
$$
\rho_{AB}^{\prime}=\left(\begin{array}{cccc}
    \frac{1-x}{4-2x} & 0 & 0 & 0 \\
    0 & \frac{1}{2+2x} & 0 & 0 \\
     0 & 0 & \frac{1}{4-2x} & 0\\
      0 & 0 & 0 & \frac{x}{2+2x}
\end{array}\right).
$$
Therefore,
$C(\rho_{AB}^{\prime},H_{AB})=\frac{1-x}{2+2x}(\epsilon^A+\epsilon^B)
+\frac{x}{4-2x}(\epsilon^A-\epsilon^B)$ and
$C(\rho_{A}^{\prime},H_{A})=2\frac{1-2x}{(2-x)(1+x)}\epsilon^A.$
Thus, $f=C(\rho_A^{\prime},H_A)-C(\rho_A,H_A) \geq 0$ for all
$x\in[0,\frac{1}{2})$, namely, the battery capacity of the subsystem
$A$ will be always enhanced, see Fig. 2.
\begin{figure}[ptb]
\includegraphics[width=0.45\textwidth]{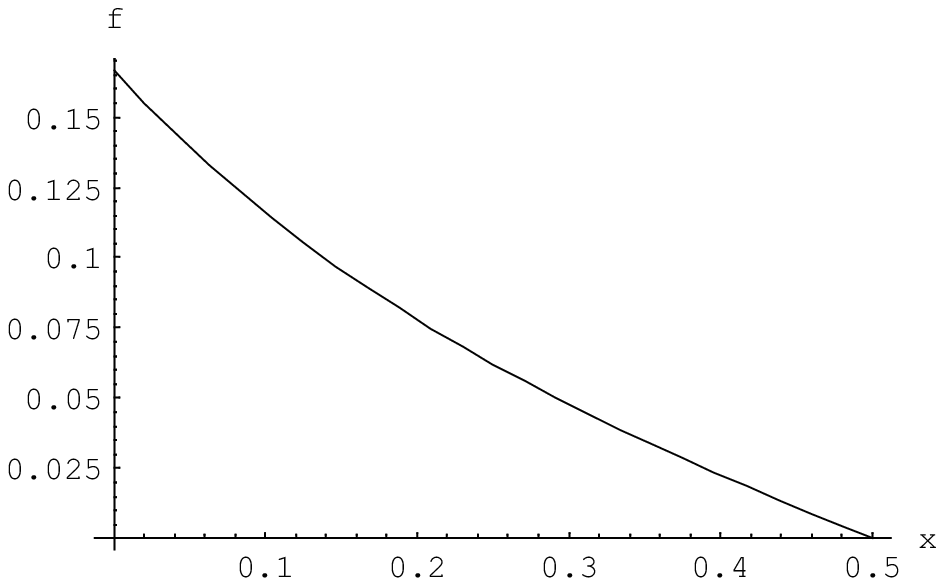}\vspace{-20.0em}\caption{The variation of the battery capacity of the sub-system A for $\rho_{AB}$ in Example 2, where $f=C(\rho_A^{\prime},H_A)-C(\rho_A,H_A)$ is a function of $x$ for $\epsilon^A=0.5$.}
\end{figure}

If we adopt Eq.(\ref{zyf3}) for $\rho_{AB}^{\prime}$, similarly we
obtain
$$
\rho_{AB}^{\prime}=\left(\begin{array}{cccc}
    \frac{\mu_0(1-x)}{2-x} & 0 & 0 & 0 \\
    0 & \frac{\mu_1}{1+x} & 0 & 0 \\
     0 & 0 & \frac{\mu_0}{2-x} & 0\\
      0 & 0 & 0 & \frac{\mu_1x}{1+x}
  \end{array}\right).
$$
Therefore, $C(\rho_{AB}^{\prime},H_{AB})=\frac{(1-x)\mu_1}{1+x}
(\epsilon^A+\epsilon^B)+\frac{x\mu_0}{2-x}(\epsilon^A-\epsilon^B)$
and $C(\rho_{A}^{\prime},H_{A})=2\frac{(\mu_1-\mu_0)x^2+2\mu_1
-(3\mu_1+\mu_0)x}{(2-x)(1+x)}\epsilon^A.$ Here, we have assumed that
$\mu_1 > \mu_0 > 2x\mu_1.$ Taking $\epsilon^A=0.5$,
$\epsilon^B=0.3$, $\mu_1=0.9$ and $\mu_0=0.1$ with $\mu_0 > 2x\mu_1$
as an example, we get $x <0.056$. It is seen from Fig. 3 that
$F=C(\rho_{AB}^{\prime},H_{AB})-C(\rho_{AB},H_{AB})$ is greater than
$0$, namely, the battery capacity of the entire system also
increases. On the other hand, as long as $x$ tends to $0$ and
$\mu_1$ tends to $1$, we have
$C(\rho_{A}^{\prime},H_{A})-C(\rho_{A},H_{A}) > 0$, i.e., the
battery capacity of the subsystem  $A$ increases too.
\begin{figure}
\includegraphics[width=0.45\textwidth]{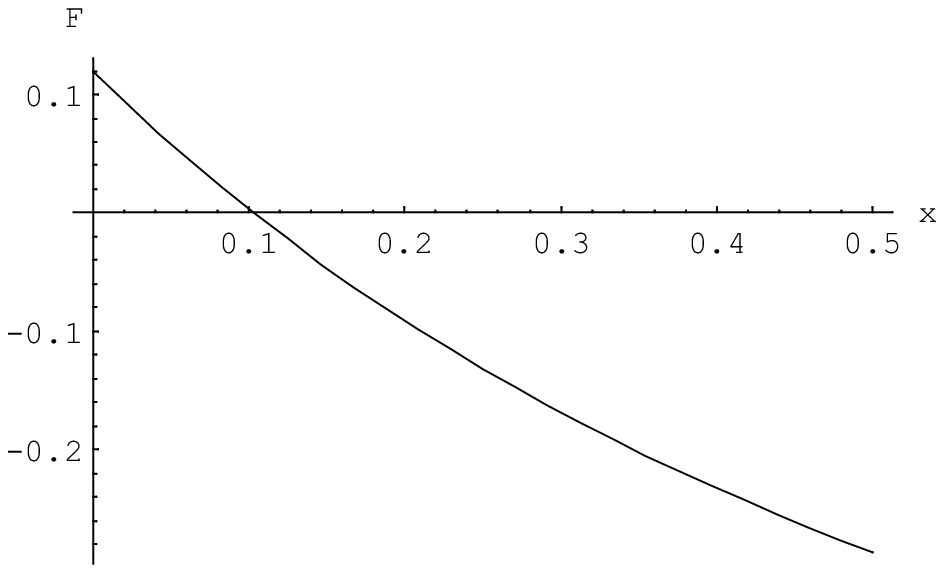}
\vspace{-20.0em}\caption{The variation of the battery capacity of the whole system for $\rho_{AB}$ in Example 2. Here, $F=C(\rho_{AB}^{\prime},H_{AB})-C(\rho_{AB},H_{AB})$
is a function of $x$ for $\epsilon^A=0.5$, $\epsilon^B=0.3$,
$\mu_1=0.9$ and $\mu_0=0.1$, where $\mu_1$ and $\mu_2$ correspond to the probabilities of choosing different measurement operators, respectively.}
\end{figure}

\section{Conclusions and Outlook}
We have studied the enhancement of the quantum battery capacities
for bipartite systems under local projective measurements. In
particular, we have investigated the cases that the initial states
are two-qubit Bell diagonal and X-type ones. We have provided
analytical results for these general initial states, which indicated
that our scheme can improve the battery capacity of the entire or
the subsystems, regardless of whether the original quantum state is
entangled or separable. There are also other studies on quantum
batteries by using quantum measurements
\cite{gjfm,gfrz,srsv2,jsjj,paku}, but with different extraction
protocols. Our study has some advantages compared with, for example, the one in the latest Ref.\cite{jsjj}. Firstly, we adopted the latest definition of quantum battery capacity, and our protocol starts directly from the bipartite quantum states, without considering the auxiliary systems. Secondly, Ref.\cite{jsjj} requires multiple measurements and the assistance of POVM measurements. Here, our protocol only involves one projective measurement. Moreover, we have provided the analytical results when the initial state is a general $2$-qubit quantum state.

The quantum battery capacity is unitary
invariant and solely given by the eigenvalues of the
quantum states and the Hamiltonian. There are
relationships between the battery capacity and the quantum resources
such as quantum entropy, entanglement and coherence \cite{hsqx,xymr}.
We have shown that the improvement of quantum battery capacity is related to both
the initial states and the combination forms of the final states. Our
results indicate that the local projection measurement
is a kind of important resource for improving the capacity of quantum
batteries. It would also interesting to investigate the
impact of general positive operator-valued measures on the improvement of
quantum battery capacities.

\bigskip
{\bf Acknowledgments:} This work is supported by the Hainan
Provincial Natural Science Foundation of China under Grant
No.121RC539; the National Natural Science Foundation of China
(NSFC) under Grant Nos. 12204137, 12075159 and 12171044; the
specific research fund of the Innovation Platform for Academicians
of Hainan Province under Grant No. YSPTZX202215 and Hainan
Academician Workstation (Changbin Yu).


\end{document}